\documentclass[letter]{aa} 

\usepackage{graphicx}
\usepackage{txfonts}
\usepackage{ulem}
\usepackage{tabularx}
\usepackage{subfigure}
\usepackage{caption}
\usepackage{hyperref}

\DeclareMathOperator*{\argmin}{argmin}

\begin{document}

        \title{Kolmogorov analysis of pulsar TOA}
        
        
        \author{N. Galikyan \inst{1,2} 
                \and
                A.A.Kocharyan\inst{3}
                \and
                V.G. Gurzadyan\inst{1,4} \fnmsep\thanks{Email: gurzadyan@yerphi.am}
        }
        
        \institute{Center for Cosmology and Astrophysics, Alikhanian National Laboratory and Yerevan State University, Yerevan, Armenia \and Department of Physics, Sapienza University of Rome, Rome, Italy
                \and School of Physics and Astronomy, Monash University, Clayton, Australia 
                \and SIA, Sapienza University of Rome, Rome, Italy
        }
        
        \date{Received XXX; accepted ZZZ}

        \abstract
        {The Kolmogorov stochasticity parameter (KSP) as a sensitive descriptor of the degree of randomness of signals was used to analyze the properties of the NANOGrav pulsar timing data associated with a stochastic gravitational wave background.  The time of arrival (TOA) data of white noise for 68 pulsars were analyzed regarding their KSP properties.  The analysis enabled us to obtain the degree of randomness of the white noise for various pulsars and to reveal its inhomogeneity, i.e., pulsars with low and high randomness of the white noise. The time dependence of the randomness in the white noise was also studied, indicating the existence of nonstationary physical processes influencing the pulsar timing. The KSP is thus acting as an indicator for the degree of the agreement between the observations and the timing models and as a test in revealing the contribution of various physical processes in the stochastic background signal. 
        }

        
        \maketitle
        
\section{Introduction}

The data on the pulsar-timing monitoring obtained by the North American Nanohertz Observatory for Gravitational Waves (NANOGrav) \citep{Ag,Ag1} and other collaborations opened a new important window to trace the variety of ongoing processes in the Universe.  The released pulsar timing array (PTA) data are linked to the stochastic gravitational-wave background (GWB) in the nanohertz band produced by binary supermassive black holes \citep{Ses,Agar,Ses1}, and they are also used to constrain several models such as those pertaining to the nature of  dark matter and the evolution of early cosmology (e.g., \citep{TG,Ben} and references therein).

The key issue in the pulsar timing is the analysis of the stochastic properties of the time of arrival (TOA) data of residuals for the studied 68 pulsars, for example the differences between the observed data and modeled parameters \citep{Lam}. Therefore, the study of tiny properties of the stochastic background by different methods and hence, of the residuals' association to white noise, can be crucial for both revealing the details from the pulsar
astrometric parameters of pulsars up those of the supermassive black hole binaries (e.g. of their orbital eccentricities) or of the contribution of other effects (see \citep{Fis} and references therein). Moreover, white noise can contain contributions from instrumental noise,  calibration of the radio telescopes, in addition to interference from terrestrial sources. Thus, while the comprehensive study of white noise is crucial for revealing the origin of the nanoherz gravitational wave background, the use of different descriptors can be of particular importance for the separation of contributions for the abovementioned effects in  white noise among others.  

For this study we applied the Kolmogorov stochasticity parameter (KSP) to analyze the properties of the white noise (cf.\citep{Bar}) of TOA associated with both the TOA data and the pulsar timing models. The Kolmogorov stochasticity parameter is a sensitive quantitative descriptor of the degree of randomness of sequences \citep{K,N}. The KSP has been applied to the analysis of various signals (e.g., \citep{N,Entr,Hof}, including human genomic sequences \citep{RS}, as well as those of an astrophysical origin (\citep{GK,KSKY,gamma,Xray,GKG}, see also \citep{G}). For example, the KSP enabled us to trace the properties of the cold spot as of a non-Gaussian region in the cosmic microwave background sky (Planck data) and to conclude on its void origin in the large-scale matter distribution \citep{CS}; the void nature of the cold spot was soon after confirmed by an galaxy survey study \citep{Sz}.  

Thus, we used the KSP for the following: (a)  to quantify how independent the TOA residuals are; (b) to reveal how  homogeneous the randomness is regarding individual pulsars, i.e., if there are pulsars attributed to low and high randomness signals; (c) to test if nonstationary effects can be traced in the white noise; and (d) to determine how closely the timing models for individual pulsars match their observations.

\section{Kolmogorov stochasticity parameter}


First, let us define the Kolmogorov stochasticity parameter as in \citep{K,N,UMN,ICTP,MMS,FA}. Let $X$ be a real-valued random variable. Its cumulative distribution function (CDF) is
\begin{equation}
    F(x) \equiv \Pr(X \le x),
    \label{eq:cdf}
\end{equation}
where $\Pr(\cdot)$ denotes the probability under the assumed probabilistic model. Throughout this section, $F$ refers to the theoretical (model/null) CDF.

Given an observed sample $x_{1},...,x_{n}$ (not necessarily distinct), let $x_{(1)}\le x_{(2)}\le\cdot\cdot\cdot\le x_{(n)}$ denote the order statistics. The corresponding empirical CDF (ECDF) is defined by
\begin{equation}
    F_{n}(x) \equiv \frac{1}{n}\sum_{i=1}^{n}\mathbf{1}\{x_{i}\le x\},
\end{equation}
which is a right-continuous step function,
\begin{equation}
    F_{n}(x) = 
    \begin{cases}
        0, & x < x_{(1)}, \\
        k/n, & x_{(k)}\le x < x_{(k+1)}, \quad k=1,\dots,n-1, \\
        1, & x_{(n)}\le x.
    \end{cases}
\end{equation}
Our basic goal is to quantify the discrepancy between the ECDF of the data and a reference (theoretical) CDF $F$. In applications, $F$ represents a null/model hypothesis (e.g., a specified distribution, or a distribution implied by a simulated noise model). When $F$ is fully specified and continuous, the resulting statistic is distribution-free under the null.

When the (one-sample) Kolmogorov distance is defined,
\begin{equation}
    D_{n} \equiv \sup_{x\in \mathbb{R}} |F_{n}(x) - F(x)|,
    \label{eq:Dn}
\end{equation}
and for its scaled form, the Kolmogorov stochasticity parameter,
\begin{equation}
    \lambda_{n} \equiv \sqrt{n}D_{n} = \sqrt{n} \sup_{x\in\mathbb{R}} |F_{n}(x) - F(x)|.
\end{equation}
Because the sample $\{X_{i}\}_{i=1}^{n}$ is random, the ECDF $F_{n}$ and thus $\lambda_{n}$ are random variables. Under the null hypothesis $H_{0}: X_{1},...,X_{n} 
{\sim} F$ (with $F$ continuous), $\lambda_{n}$ has a distribution function
\begin{equation}
    \Phi_{n}(\lambda) \equiv \Pr(\lambda_{n}\le\lambda),
\end{equation}
which converges to a universal limiting distribution $\Phi(\cdot)$ independent of $F$:
\begin{equation}
\Phi(\lambda) = \sum_{k=-\infty}^{+\infty}{(-1)^k e^{-2k^2\lambda^2}},\\
\Phi(0) = 0.
\label{eq:phi}
\end{equation}
This is the statement of Kolmogorov's theorem \citep{K}.


Note that, according to this theorem and the property of the limiting function $\Phi$, the probable values of $\lambda$ are within 0.3 and 2.2.\ In other words, at its lower values,  $\Phi(\lambda)$ tends rapidly to 0 and to 1 at higher values.

An important feature of the KSP criterion is its applicability to relatively small sequences (orbits of dynamical systems) as is explicitly shown in \citep{N}, which compared two sequences of 15 two-digit numbers and concluded that the stochasticity probability of the one sequence is approximately 300 times higher than for other one. Then, for random sequences, the stochasticity parameter $\lambda_n$ has a distribution close to the function $\Phi(\lambda)$, while the distribution is different from that function for nonrandom sequences.

The numerical experiments performed in \citep{G2011}, using
a single scaling of the ratio of stochastic to regular components of the generated sequences including a uniform distribution, revealed the efficiency of the KSP as an indicator of the degree of randomness of composite signals.

\section{Pulsar timing}
Pulsar timing is based on measuring TOAs of radio pulses with high precision and comparing them to the predictions of a timing model. The timing model encodes all deterministic contributions to the TOAs—such as the pulsar’s spin evolution, astrometric parameters, binary motion (when present), and propagation effects. Any mismatch between the measured TOAs and the timing model—i.e., the timing residuals—contains information about additional physical or instrumental processes not captured by the deterministic model.

These stochastic contributions are commonly separated into low-frequency, temporally correlated “red noise’’ (e.g., intrinsic spin noise or a stochastic gravitational-wave background), which are the type of signals that researchers usually look for. It is important to note that not all pulsars exhibit a red-noise component, and the contribution to the red noise can come from various effects \citep{Fis}, which can
introduce characteristic structures in timing residuals. High-frequency, uncorrelated “white noise’’ mainly arises from radiometer noise and measurement uncertainties, representing random fluctuations in the time series. The white-noise component can be calculated as $WhiteNoise = TOA - TimingModel - RedNoise \ (if present)$. 

For our analysis, we used PINT \citep{PINT,PINT1,PINT2}, a high-precision open-source package for pulsar timing data analysis written in Python. It is developed in collaboration with NANOGrav and is largely used for analysis \citep{Ag,Ag9,Smith}, among other things. It allows one to read TOA data and load timing models with their parameters, for example, provided by NANOGrav \citep{Ag1}. Moreover, in PINT, one can simulate the white and red noise components of pulsar signals separately and estimate the white-noise contribution. The simulated white noise and the observed white noise should have the same statistical properties, i.e., the signal that we refer to as white noise in the observed data is indeed a white-noise signal. To test this, we used the KSP as a statistical characteristic of a signal and compared the observed signal with the simulated one as described in \ref{sec:beta_const}.

\section{Analysis}

Consider a time series of length $N$ that represents the residual component of the TOAs \citep{DATA}. To obtain the KSP characteristic histogram for the series, we calculated $\lambda$ inside a sliding window of length $10^{-1}\cdot N$ with a sliding step of $10^{-3}\cdot N$. Inside the window, we randomly selected $80\%$ of the points and calculated $\lambda$ using the generalized normal distribution as the theoretical distribution. Its probability density is given via~Eq.(\ref{eq:teor}). The final $\lambda$ for that window is the mean of the $80\%$s which was repeated five times. Namely, the algorithm estimated the distribution of $\lambda$, and its running allowed the $\lambda$ distribution's stability to be estimated four more times, as $80\%$ of the subsamplings each yielded different distributions. This enabled us to ensure the stability of the procedure, to define the errors bars, and to decrease the influence of outliers, if any.  The specific choice of parameters -- the $80\%$ subsampling and the $5\times5$ repetitions -- were made in order to adequately represent the distributions for comparison, while avoiding both over-averaging and undersampling:

\begin{equation}\label{eq:teor}
    \begin{aligned}
            &f_\beta(x) = \frac{1}{2\alpha\Gamma(1/\beta)}e^{\left(\frac{|x-\mu|}{\alpha}\right)^\beta} \\
            &\alpha = \sqrt{\frac{\Gamma(1/\beta)}{\Gamma(3/\beta)}}; \quad \mu=0
    \end{aligned}
.\end{equation}

\begin{figure}[h!]
\centering

\begin{minipage}{\columnwidth}
    \centering
    \includegraphics[width=\columnwidth]{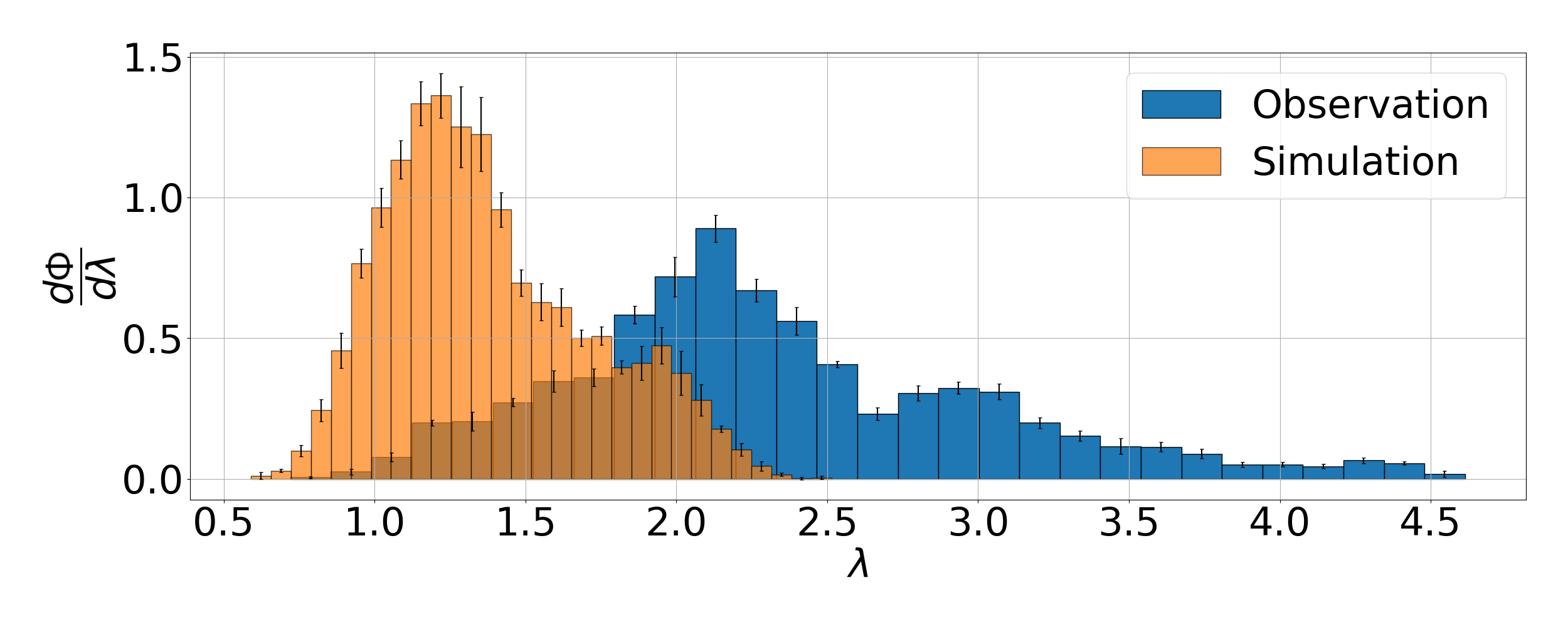}
  (a) J1022+1001
\end{minipage}

\begin{minipage}{\columnwidth}
  \centering
  \includegraphics[width=\columnwidth]{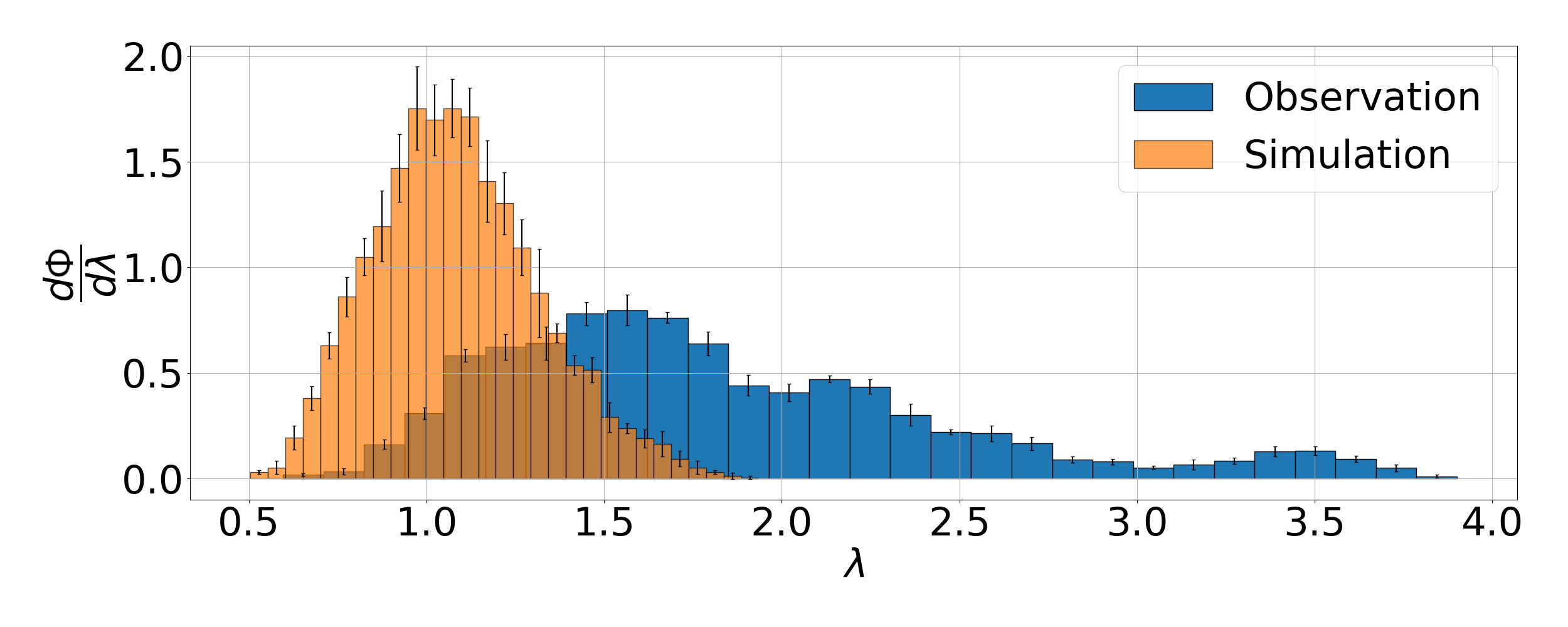}
  (b) J1125+7819
\end{minipage}

\begin{minipage}{\columnwidth}
  \centering
  \includegraphics[width=\columnwidth]{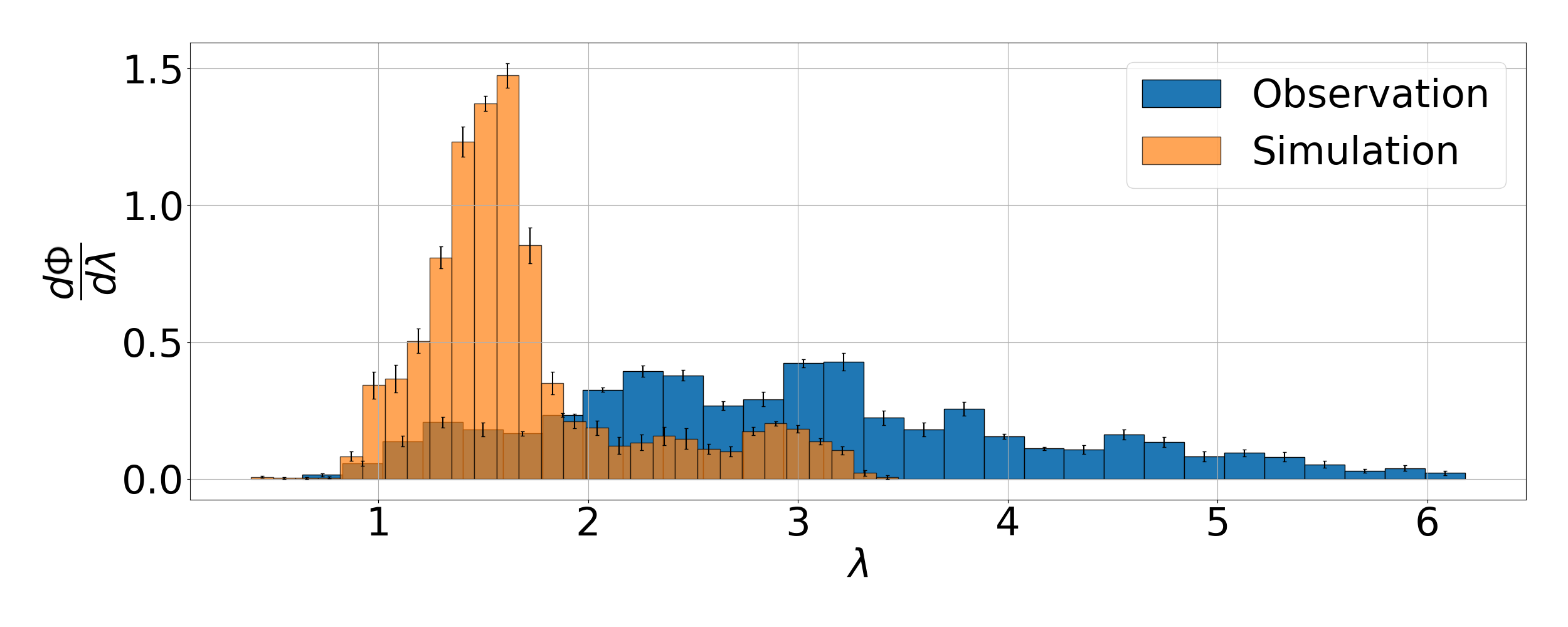}
  (c) J1903+0327
\end{minipage}

\caption{KSP distributions for the simulated/observed data of pulsars' white noise, when the histograms do differ. Orange and blue histograms correspond to simulations and observations, respectively.}
\label{fig:different}

\end{figure}

\begin{figure}[h!]
\centering

\begin{minipage}{\columnwidth}
  \centering
  \includegraphics[width=\columnwidth]{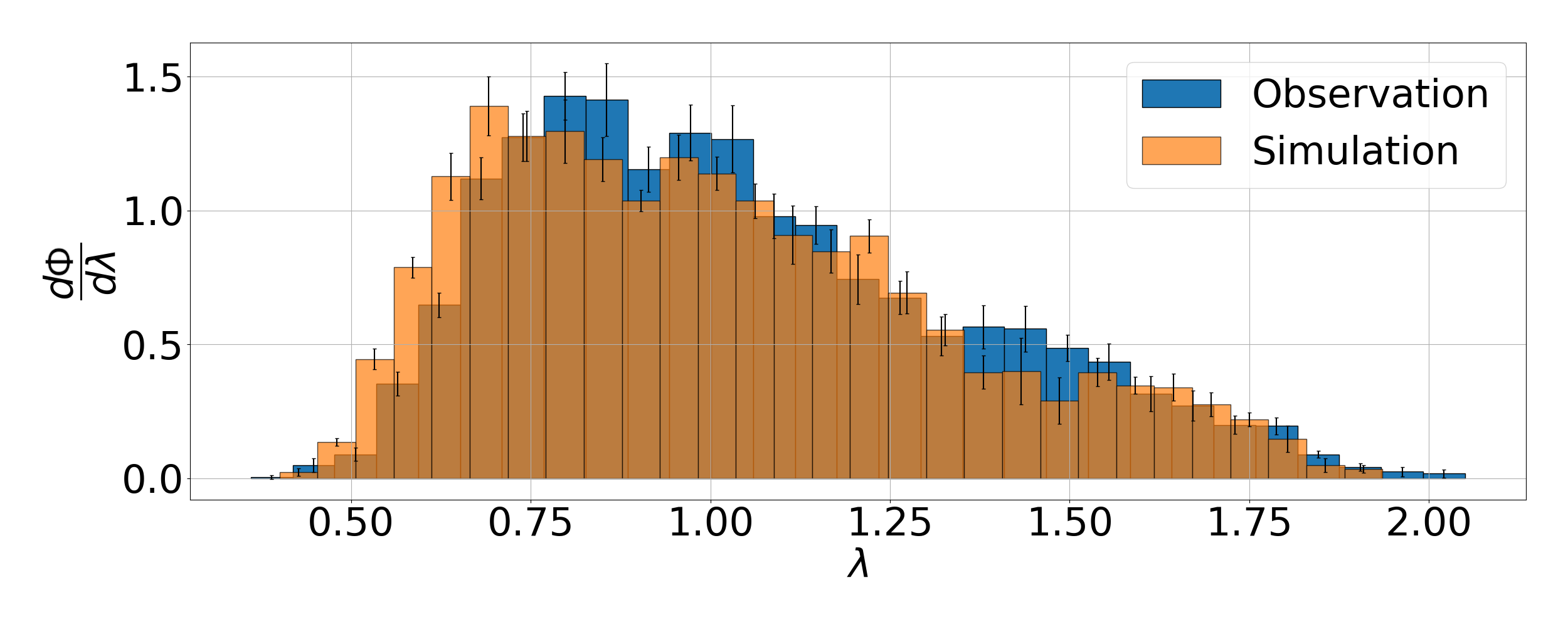}
  (a) J1745+1017
\end{minipage}

\begin{minipage}{\columnwidth}
  \centering
  \includegraphics[width=\columnwidth]{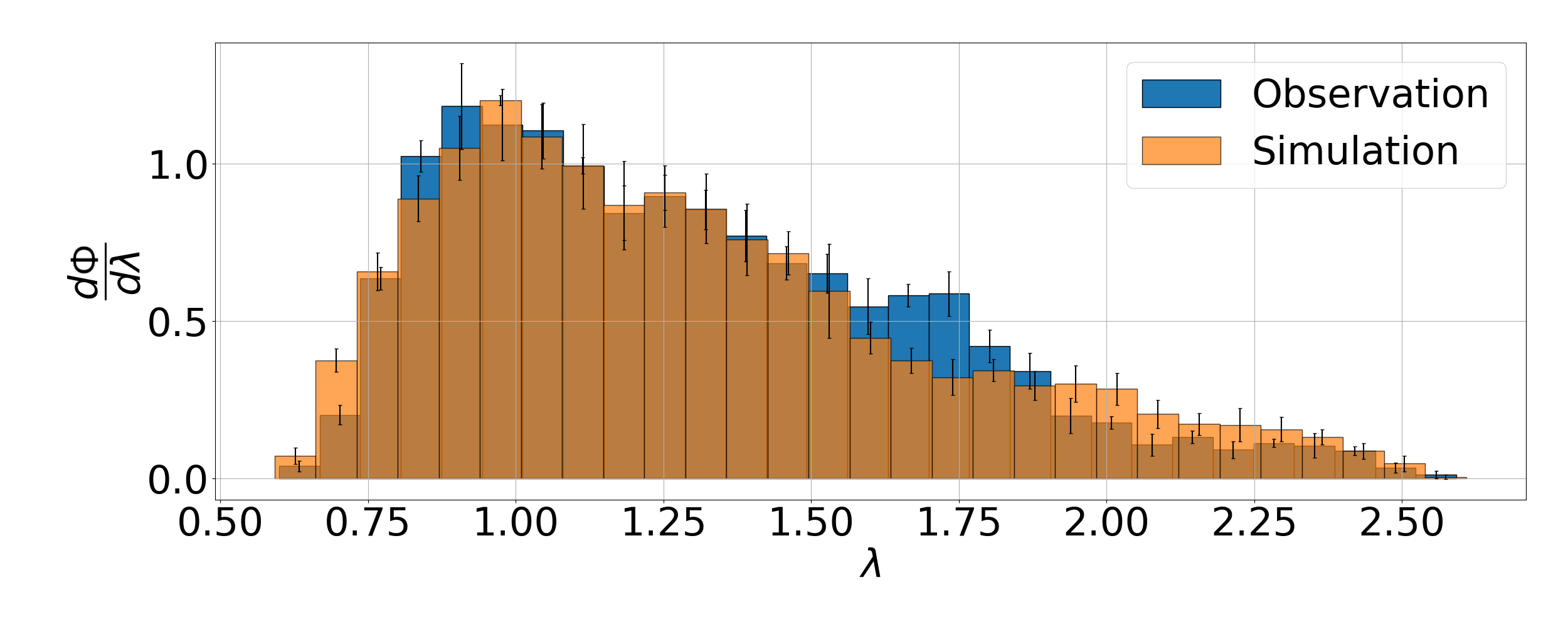}
  (b) J1910+1256
\end{minipage}

\caption{Same as in Figure \ref{fig:different}, but for sample pulsars when the KSP distributions for the simulated/observed data are similar.}
\label{fig:same}
\end{figure}

In (\ref{eq:teor}) we have only a single free parameter $\beta$ that describes the sharpness of the distribution. The scale parameter $\alpha$ was chosen such that the theoretical distribution had a unit variance. We considered two scenarios of data normalization and a choice of $\beta$. First, when $\beta$ was a global parameter of a time series as described in \ref{sec:beta_const}, and, second, where $\beta$ was local and had a time dependence (\ref{sec:beta_time}). 

\subsection{$\beta=\mathrm{const}$}\label{sec:beta_const}

\begin{figure}[h!]
    \centering
    \includegraphics[width=\columnwidth]{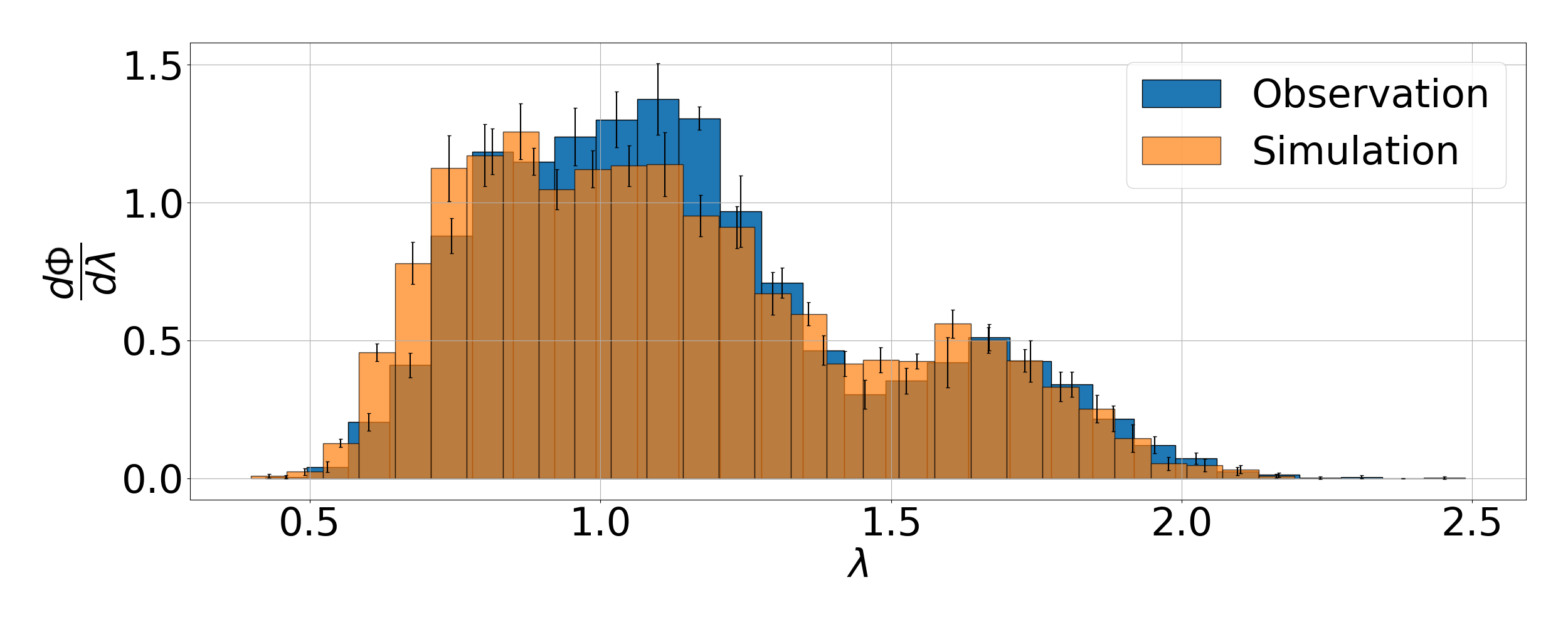}
    \caption{Case of pulsar J1125+7819, when the KSP distributions became similar after introducing $\beta(t)$.}
    \label{fig:changed}
\end{figure}

\begin{figure}[h!]
    \centering
    \includegraphics[width=\columnwidth]{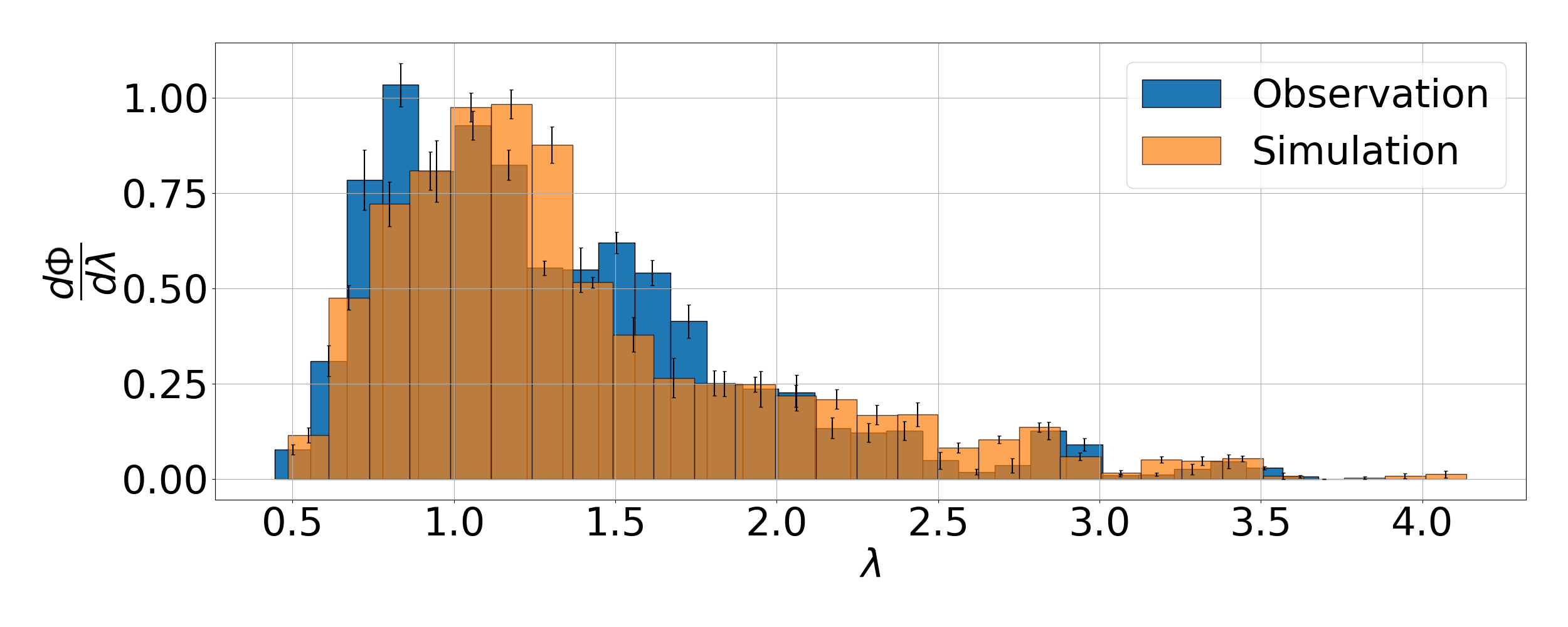}
    
    \caption{ KSP distributions for pulsar J1910+1256. Introducing the time dependence in $\beta$ does not significantly affect $\chi^2$.}
    \label{fig:samet}
\end{figure}

\begin{figure}[h!]
    \centering
    \includegraphics[width=\columnwidth]{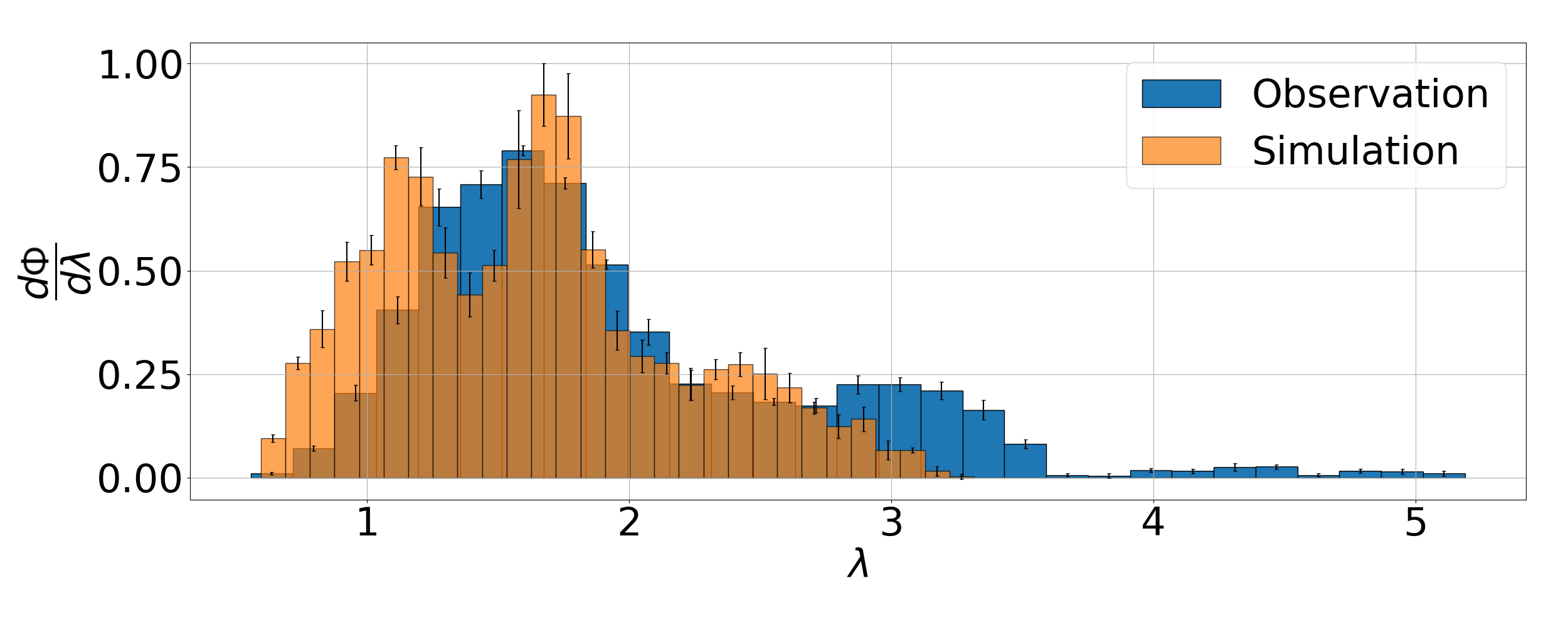}
    \caption{KSP distributions for the pulsar J1022+1001 after introducing the time dependence in $\beta$. Although $\langle\chi^2\rangle$ is smaller in the $\beta(t)$ approach (Tab. \ref{tab:chi_betat}) than in the $\beta=\mathrm{const}$ case (Tab. \ref{tab:chi}), they are of the same order.}
    \label{fig:differentt}
\end{figure}

Consider a time series of residuals of a pulsar, $\{R_i\}_{i=1}^{N}$, where $N$ is the length of the series. We assumed that they are identically distributed. In this scenario, we first normalized our series by subtracting the mean of the series $\overline{R}$ and dividing by its standard deviation $\Delta R$. Hence, we obtain normalized series $r_i = \frac{R_i-\overline{R^j}}{\Delta R}$.

We performed this procedure on observed and simulated data using PINT \citep{PINT,PINT1,PINT2}. Simulations used the same models, and the parameters within them were used to extract residuals from the TOA. We observed that some of the pulsars' KSP distributions are not similar, as shown in Fig. \ref{fig:different}, while for others they are, as shown in Fig. \ref{fig:same}.

For a quantitative picture, we created five epochs of simulations for each pulsar presented, and calculated the reduced $\chi^2$ of the KSP histograms between observed and simulated data. To calculate $\chi^2$, we linearly interpolated histogram values and their corresponding errors into the same bins. Table \ref{tab:chi} shows the minimum, maximum, mean, median, and standard deviation of $\chi^2$ for each pulsar. Thus, the key point of our analysis is that the same numerical scheme was performed both for the data of observations and simulations.

\subsection{$\beta=\beta(t)$}\label{sec:beta_time}

Now we assume that the properties of the distribution may change over time, hence $\beta$ has a time dependence. We calculated $\beta$ around each residual using maximum likelihood estimation. The algorithm for finding $\beta_i$ for a residual time series $R_i$ is the following:
\begin{enumerate}
    \item We weighed each residual value by weight,
    \[
        w_{ik} \propto \exp\left[-\frac{1}{2} \left(\frac{t_k - t_i}{h}\right)^2\right], \qquad k= \overline{1;N,}
    \]
    where $t_i$, $t_k$ are the time moments for $R_i$ and $R_k$, respectively, and $h$ is a hyperparameter that was chosen via cross-validation.
    \item We calculated the weighted mean and standard deviation 
    and obtained normalized time series for each time moment $t_i$\textbf{:} $r_{ik} = \frac{R_k - \overline{R_i}}{\Delta R_i}$.
    \item We estimated $\beta_i$ as
    \[
    \beta_i = \argmin_\beta \left[ -\sum_k w_{ik}\log f_\beta(r_{ik})\right].
    \]
\end{enumerate}

\begin{table}[h!]
    \centering
    \caption{Reduced-$\chi^2$ data for each pulsar for $\beta = const$.}
    \begin{tabularx}{\columnwidth}{|c|X|X|X|X|X|}
        \hline
         & min($\chi^2$) & max($\chi^2$) & med($\chi^2$) & $\langle\chi^2\rangle$ & $\sigma(\chi^2)$ \\
        \hline
        J1022+1001 & 64 & 110 & 102 & 92 & 18 \\
        \hline
        J1125+7819  & 40  & 150 & 70 & 80 & 40 \\
        \hline
        J1745+1017 & 2.7 & 7.8 & 4.4 & 4.5 & 1.8 \\
        \hline
        J1802-2124  & 52  & 86  & 72 & 71 & 12 \\
        \hline
        J1832-0836 & 2 & 8 & 5 & 5 & 2 \\
        \hline
        J1903+0327 & 68 & 112 & 78 & 84 & 15 \\
        \hline        
        J1910+1256 & 2 & 20 & 5 & 8 & 7 \\
        \hline
        J1918-0642 & 5.3 & 7.6 & 6.4 & 6.4 & 0.8 \\
        \hline
        J2043+1711 & 4 & 17 & 6 & 8 & 5 \\
        \hline
        J2317+1439 & 36 & 73 & 48 & 51 & 14 \\
        \hline
    \end{tabularx}
    \label{tab:chi}
\end{table}

The following procedure of calculating the KSP histogram is the same, with the exception that we normalized the distribution inside the sliding window similarly to Section \ref{sec:beta_const}, i.e., without the weights. Table \ref{tab:chi_betat} shows the $\chi^2$ values for this scenario.

\begin{table}[h!]
    \centering
        \caption{Same as in Table 1, but for $\beta = \beta(t)$.}       
    \begin{tabularx}{\columnwidth}{|c|X|X|X|X|X|}
        \hline
        & min($\chi^2$) & max($\chi^2$) & med($\chi^2$) & $\langle\chi^2\rangle$ & $\sigma(\chi^2)$ \\
        \hline
        J1022+1001 & 21 & 60 & 32 & 36 & 13 \\
        \hline
        J1125+7819 & 2  & 14 & 8  & 7  & 4  \\
        \hline
        J1745+1017 & 5 & 30 & 11 & 13 & 8 \\
        \hline
        J1802-2124 & 30 & 90 & 60 & 60 & 20 \\
        \hline
        J1832-0836 & 2 & 9  & 5 & 5 & 2 \\
        \hline
        J1903+0327 & 18 & 54 & 43 & 41 & 12 \\
        \hline
        J1910+1256 & 5 & 19 & 7 & 10 & 5 \\
        \hline
        J1918-0642 & 6  & 23 & 16  & 15  & 6 \\
        \hline
        J2043+1711 & 5  & 15 & 11  & 10  & 3 \\
        \hline
        J2317+1439 & 12  & 26 & 14 & 18 & 6 \\
        \hline
    \end{tabularx}
    \label{tab:chi_betat}
\end{table}

Figures (\ref{fig:changed}, \ref{fig:samet}, \ref{fig:differentt}) illustrate examples of three differentiable cases: non-similar distributions became similar; and similar and non-similar KSP distributions remained as they were.

\section{Conclusions}

We analyzed NANOGrav pulsar timing data using Kolmogorov’s method. Namely, we concentrated on the white-noise component of the observations, as it is the remainder after subtracting the known physical contributions from the TOAs. Comparing the observed white noise with simulated white noise and using the Kolmogorov stochasticity parameter of the signal as a metric enabled us to probe the possible physical components in the TOAs. These physical components may be deterministic, and thus they might be included in the timing models, or stochastic, as part of the red noise (e.g., the GWB). The analysis was performed blindly with the same algorithm applied to all pulsars considered.

Assuming a generalized normal distribution — with a single free $\beta$ parameter describing the sharpness of the distribution — as the underlying distribution for the white noise, we find both cases, when the white noise stochasticity agrees with the observations and simulations and cases where it does not. The introduction of a time-dependent $\beta(t)$ improved the similarity; however, instances remain in which the observational and simulated stochasticity still disagree. This can indicate the existence of nonstationary physical processes influencing the pulsar timing.

The KSP method can be used to test different models affecting pulsar timing, including for example modified theories of gravity or models for the dark sector of the universe (e.g., \citep{GFC1,GFC2,GFC3}). Namely, then the task will be to reproduce TOAs with the observed KSP properties, especially depending upon the availability of higher precision and more observational data.

\begin{acknowledgements}
We are thankful to the referee for many valuable comments. We acknowledge the use of The NANOGrav 15-Year Data Set; https://zenodo.org/records/16051178 and ASCL Code Record of  Astrophysics Source Code Library ASCL.net; ascl:1902.007.
\end{acknowledgements}




\label{lastpage}
\end{document}